\documentclass[aps,prb,twocolumn]{revtex4} 

\usepackage{graphicx}
\usepackage{dcolumn}
\usepackage{bm}
\usepackage{amsmath} 

\newcommand{\comment}[1]{}

\begin{document}

\title{Gaussian Reformulation of the Feynman Path Integral
for Quantum Statistical Mechanics
with Results for the Second Virial Coefficient of $^4$He}


\author{Phil Attard}
\affiliation{ {\tt phil.attard1@gmail.com}  June 13--July 12 2026}


\begin{abstract}
The Feynman path integral for quantum statistical mechanics
is reformulated as Gaussian sampling of the neighborhood
of each position configuration.
The variance and mean are obtained
from ring polymer statistics on a lattice,
and from the high temperature expansion
of the Wigner-Kirkwood commutation function, respectively.
The algorithm avoids multiple temperature nodes for each configuration
and the need for numerical cancelation in the statistical averages,
which are problematic
for conventional path integral quantum Monte Carlo.
Analytic and simulation results for the second virial coefficient
of helium are compared to laboratory measurements.
\end{abstract}


\maketitle

%
\section{Introduction}
\setcounter{equation}{0} \setcounter{subsubsection}{0}
\renewcommand{\theequation}{\arabic{section}.\arabic{equation}}
%

The challenge posed by quantum statistical mechanics
arises from the Maxwell-Boltzmann operator,
whose trace is required.
This has as exponent the sum of the kinetic energy
and the potential energy operators,
which do not commute.
For any realistic system the energy eigenfunctions and eigenvalues
are unknown
and it is not feasible to calculate their spectrum numerically.

One computer simulation approach that has been successful
is based on the Feynman path integral,
which factorizes each position configuration
into a large number of configurations
each at reduced temperature and connected by Gaussian bonds
to form a ring polymer
(Allen and Tildesley 1987, Ceperley 1995, Feynman and Hibbs 1965).
A second computer approach is based upon various high temperature expansions
of the so-called Wigner-Kirkwood commutation function
(Allen and Tildesley 1987, Kirkwood 1933, Wigner 1932),
which can be formulated as the Maxwell-Boltzmann operator
acting on the momentum eigenfunctions (Attard 2018b, 2021).

A significant computational problem with the path integral approach
is that it requires $M$ replicas of each position configuration,
each of which has to be independently simulated
(Allen and Tildesley 1987, Ceperley 1995).
For liquid helium at low temperatures,
$M = {\cal O}(10^3)$ in the so-called primitive approximation
(Ceperley 1995).
More sophisticated approximations reduce
this to $M = {\cal O}(50)$,
albeit with additional computer cost for each one.
This has limited the size of the simulations to $N=64$ atoms (Ceperley 1995),
compared to classical Monte Carlo that typically has
$N=10^3$--$10^4$ atoms (Allen and Tildesley 1987, Attard 2002, 2012).
A second problem with the path integral approach
is that the averages of thermodynamic quantities
include contributions from the Gaussian bonds
that connect adjacent configurations around the ring polymer path.
These are not physical quantities
and so they must cancel numerically in the average,
which is problematic as their individual magnitudes increase
with the length of the path $M$.

The high temperature expansions of the Wigner-Kirkwood commutation function
that have been implemented have their own challenges (Attard 2018b, 2021).
The series they produce are tedious to derive,
with the terms becoming increasingly complicated
with increasing numbers of increasingly higher order gradients
of the interaction potential.
These gradients pose problems in the core region of the atomic interaction,
and the series themselves can be slow to converge at lower temperatures.
The two main advantages of the high temperature expansions
over path integral Monte Carlo are
that the size of the system
that can be simulated is similar to classical simulations,
$N=10^3$--$10^4$,
and that wave function symmetrization is treated  with greater sophistication
(Attard 2018b, 2021, 2025a).

This paper offers an alternative to these two approaches
that is in a sense their hybrid.
The point is made that one interpretation of the ring polymer
produced by the Feynman path integral
is that it is just a means of sampling configurations
in the vicinity of the original position configuration of interest.
There is no intrinsic need
for the bonds between the configurations
or the order in which they lie on the path,
as indeed these have to be canceled from the final averages.
The length of the path $M$ is not a physical quantity
as it has to be taken to infinity,
which suggests that it is the points visited by the path
rather than the path itself that are important.
If the correct sampling density of the ring polymer can be deduced,
then the need to generate the ring polymer
or to follow the Feynman path is obviated,
thus avoiding the two problems mentioned above.

The physical motivation for this interpretation
lies in the Heisenberg uncertainty relation.
Roughly speaking we may say that this creates a certain
fuzziness for each position configuration,
so that the state of the system is determined
by the neighborhood rather than by the position point itself.
The well-known zero point energy
is a related phenomenon that shows that the system cannot be pinned down
precisely to the minimum in the potential energy.
Of course, the non-commutativity
of the kinetic and potential energy operators
in the exponent of the Maxwell-Boltzmann operator
is itself closely related to the Heisenberg uncertainty relation.

The paper is set out as follows.
Section~\ref{Sec:Analysis} reprises the fundamental formulation
of quantum statistical mechanics,
beginning with the Wigner-Kirkwood commutation function
for the weight of classical phase space.
From this the Feynman path integral is derived
in both classical phase space
and in the more conventional position configuration space.
Expressions are given for the virial pressure and the energy,
where the dependence on path length $M$
and the need for the numerical cancelation of large terms is made explicit.
Section~\ref{Sec:Gauss} contains the new results
as it gives the reformulation
of the path integral as a Gaussian probability.
Section~\ref{Sec:B2} derives an analytic expression
for the second virial coefficient
based on the leading high temperature term
in the exact expansion for quantum statistical mechanics
in classical phase space.
A quantum Monte Carlo algorithm for the Gaussian reformulation
of the path integral is given in \S~\ref{Sec:Results}.
In that section simulation results
for the second virial coefficient for $^4$He
are compared with laboratory measurements
and with the analytic results.

%
\section{Fundamentals} \label{Sec:Analysis}
\setcounter{equation}{0} \setcounter{subsubsection}{0}
\renewcommand{\theequation}{\arabic{section}.\arabic{equation}}
%

\subsection{Phase Space Weight}

For a subsystem of  $N$ identical bosons,
a point in classical phase space is
${\bf \Gamma} = \{{\bf q},{\bf p}\}$.
Here the position configuration is
${\bf q} = \{ {\bf q}_1,{\bf q}_2,\ldots, {\bf q}_N\}$
and the momentum configuration is
${\bf p} = \{ {\bf p}_1,{\bf p}_2,\ldots, {\bf p}_N\}$,
where the position of boson $j$ is
${\bf q}_j = \{ {q}_{jx},{q}_{jy},{q}_{jz}\}$,
and its momentum is
${\bf p}_j = \{ {p}_{jx},{p}_{jy},{p}_{jz}\}$.
In many cases the momentum may belong to the continuum;
but for Bose-Einstein condensation and superfluidity quantized momentum
appears to be necessary (Attard 2025a, 2025b, 2025g).

For a canonical equilibrium system of temperature $T$ and volume $V$,
the phase space probability density is
(Attard 2018b, 2021)
\begin{equation}
\wp({\bf \Gamma}) =
\frac{
e^{-\beta {\cal H}({\bf \Gamma})}
e^{W({\bf \Gamma})} \eta({\bf \Gamma})
 }{ N! h^{3N}Z },
\end{equation}
where Planck's constant is $h=6.63\times10^{-34}$\,J\,s
and the inverse temperature is $\beta = 1/k_{\rm B}T$,
$k_{\rm B} = 1.38 \times 10^{-23}$\,J/K
being Boltzmann's constant.
The partition function, $Z(N,V,T)$,
normalizes the probability and its logarithm gives the total entropy.
Also the classical Hamiltonian is
${\cal H}({\bf \Gamma}) = {\cal K}({\bf p}) + U({\bf q})$,
where ${\cal K}({\bf p}) = p^2/2m = \sum_{j=1}^N p_j^2/2m$
is the kinetic energy,
$m$ being the mass of a boson.
Usually the potential energy consists of central pair potentials,
$U({\bf q}) = \sum_{j<k}^N u(q_{jk})$;
one can write $u_{jk}  \equiv u(q_{jk})$.


The Wigner-Kirkwood (Kirkwood 1933, Wigner 1932)
commutation function $W({\bf \Gamma})$ is defined by
(Attard 2018b, 2021)
\begin{equation} \label{Eq:Wdefn}
e^{-\beta {\cal H}({\bf \Gamma})}
e^{W({\bf \Gamma})}
=
e^{{\bf p}\cdot{\bf q}/{\rm i} \hbar}
e^{-\beta \hat{\cal H}({\bf q})}
e^{-{\bf p}\cdot{\bf q}/{\rm i} \hbar} .
\end{equation}
In previous notation
this corresponds to $\omega_p =e^{W_p}$ (Attard 2018b Eq.~(2.6)).
For the quantum statistical weight one can use either $\omega_p \eta_q$
or else $\omega_q \eta_p$,
with $\omega_q^* = \omega_p$ and $\eta_q^* = \eta_p$.

On the right hand side of the definition of the commutation function
the Hamiltonian operator appears,
$ \hat{\cal H}({\bf q}) =  \hat{\cal K}({\bf q}) +  U({\bf q})$.
The Fourier factors are unnormalized, unsymmetrized momentum eigenfunctions.
The non-zero commutator $[\hat{\cal K}, U] \ne 0$
is what makes $W \ne 0$,
and so it reflects the Heisenberg uncertainty relation.



The symmetrization function
is the ratio of unpermuted to permuted momentum eigenfunctions
summed over all permutations  (Attard 2018b, 2021),
\begin{equation}
\eta({\bf \Gamma})
=
\sum_{\hat{\rm P}}
e^{-[{\bf p}-\hat{\rm P}{\bf p}]\cdot{\bf q}/{\rm i}\hbar} .
\end{equation}
This is for bosons and it corresponds to $\eta_q$  (Attard 2018b Eq.~(2.4)).
\comment{ 
That part of the grand potential due to symmetrization
is given by the average of this,
\begin{eqnarray}
e^{-\beta \Omega_{\rm sym}}
& = &
\langle \eta({\bf \Gamma}) \rangle_W
\nonumber \\ & = &
e^{ \langle \stackrel{\circ}{\eta}({\bf \Gamma}) \rangle_W } .
\end{eqnarray}
In the second equality,
which is believed to be exact in the thermodynamic limit
(Attard 2018b \S~III\,B\,1),
the sum over single permutation loops is
$\stackrel{\circ}{\eta}\!\!({\bf \Gamma}) =
\sum_{l=2}^\infty \eta^{(l)}({\bf \Gamma})$,
where the $l$-loop symmetrization function is
\begin{equation}
\eta^{(l)}({\bf \Gamma})
=
\sum_{j_1,\ldots,j_l}^N\!\!\!'\; \prod_{k=1}^{l}
e^{ - {\bf p}_{j_k} \cdot {\bf q}_{j_k,j_{k+1}} /{\rm i}\hbar}
 , \quad j_{l+1} \equiv j_1 .
\end{equation}
The sum is over the unique directed cyclic permutations
of all subsets of $l$-bosons.
The $l$-loop grand potential is
(Attard 2018b, 2021)
\begin{equation}
-\beta \Omega_W^{(l)}
=
\langle \eta^{(l)}(\Gamma) \rangle_W .
\end{equation}
(The monomer grand potential is just the logarithm
of the partition function in the absence of the symmetrization function.)
Several tricks have been found to facilitate
the computation of the  $l$-loop symmetrization function
(Attard 2021 \S5.4.2).
On the high temperature side of the $\lambda$-transition,
we expect that only position loops with consecutive particles
separated by less than about the thermal wavelength will contribute.

} 

\subsection{Path Integral}

The Feynman path integral method  is justified by
the Trotter factorization (Ceperley 1995, Feynman and Hibbs 1965,
Trotter 1959)
\begin{equation}
e^{-\tau (\hat {\cal K} + U) + \tau^2[ \hat {\cal K} , U]/2}
=
e^{-\tau\hat {\cal K}} e^{-\tau U} .
\end{equation}
This is exact.
The order of the factors on the right hand side
is determined by the commutation function on the left hand side.
We take $\tau = \beta /M$, with $M$ the number of temperature points.
In the limit $M \to \infty$, the commutator may be neglected,
which leaves the Maxwell-Boltzmann operator on the left hand side.

The momentum eigenfunctions,
$\phi_{{\bf p}}({\bf q}) = V^{-N/2} e^{-{\bf p}\cdot{\bf q}/{\rm i}\hbar}$,
form an  orthogonal  complete set
\begin{equation}
\langle \phi_{{\bf p}'} | \phi_{{\bf p}''} \rangle
=
\int_V {\rm d}{\bf q}\;
\phi_{{\bf p}'}({\bf q})^*  \,\phi_{{\bf p}''}({\bf q})
= \delta_{{\bf p}',{\bf p}''}
\end{equation}
and
\begin{equation}
\sum_{\bf p} \phi_{\bf p}({\bf q}')^* \phi_{\bf p}({\bf q}'')
= \delta({\bf q}'-{\bf q}'') .
\end{equation}
Notice that these rely upon quantized momentum states,
whose spacing is $\Delta_p = 2\pi\hbar/L$,
and integrals over the cubic volume $V=L^3$.
It is usual to take the continuum momentum limit,
which implies $ V \to \infty$.
However,
below we derive the expression for the virial pressure.
Since the volume derivative of the partition function
is usually accomplished by scaling the position configuration by $L$,
we take care to use the quantized momenta for the derivation,
and we defer the continuum limit until the end.
(Actually there is no real need for this pedantry.)

Applying the Trotter formula with $\tau \to 0$
to the Maxwell-Boltzmann operator 
acting on an arbitrary function of the position configuration,
the orthogonality and  completeness of the momentum eigenfunctions
give
\begin{eqnarray} \label{Eq:eHf-disc}
\lefteqn{
e^{-\tau \hat{\cal H}({\bf q})} f({\bf q})
}  \\
& = &
e^{-\tau \hat{\cal K}({\bf q})}
e^{-\tau U({\bf q})} f({\bf q})
\nonumber \\ & = &
\int_V {\rm d}{\bf q}'\;
\delta({\bf q}'-{\bf q})
e^{-\tau \hat{\cal K}({\bf q}')}
e^{-\tau U({\bf q}')} f({\bf q}')
\nonumber \\ & = &
\frac{1}{V^N} \sum_{ {\bf p}'} \int_V {\rm d}{\bf q}'\;
e^{{\bf p}'\cdot{\bf q}/{\rm i}\hbar}
e^{-{\bf p}'\cdot {\bf q}'/{\rm i}\hbar}
e^{-\tau \hat{\cal K}({\bf q}')}
e^{-\tau U({\bf q}')} f({\bf q}')
\nonumber \\ & = &
\frac{1}{V^N} \sum_{ {\bf p}'} \int_V {\rm d}{\bf q}'\;
e^{-{\bf p}'\cdot[{\bf q}'-{\bf q}]/{\rm i}\hbar}
e^{-\tau {\cal K}({\bf p}')}
e^{-\tau U({\bf q}')} f({\bf q}') .\nonumber
\end{eqnarray}
The final equality uses the Hermitian nature
of the kinetic energy operator,
which allows it to act to the left because of the integral over ${\bf q}'$.
Taking the continuum limit of this,
$\sum_{ {\bf p}'} \Rightarrow \Delta_p^{-3N} \int {\rm d}{\bf p}'$,
and completing the square of the kinetic energy, gives
\begin{eqnarray} \label{Eq:eHf-cont}
\lefteqn{
e^{-\tau \hat{\cal H}({\bf q})} f({\bf q})
} \nonumber \\
& = &
\frac{1}{h^{3N}}
\int {\rm d}{\bf p}' \!
\int {\rm d}{\bf q}'\;
e^{-{\bf p}'\cdot[{\bf q}'-{\bf q}]/{\rm i}\hbar}
e^{-\tau {\cal K}({\bf p}')}
e^{-\tau U({\bf q}')} f({\bf q}')
\nonumber \\ & = &
\Lambda_\tau^{-3N}
\int {\rm d}{\bf q}'\; e^{-\pi [{\bf q}-{\bf q}']^2/\Lambda_\tau^2}
e^{-\tau U({\bf q}')} f({\bf q}') .
\end{eqnarray}
Here and below the thermal wavelength is
$\Lambda_\tau = \sqrt{2\pi\hbar^2 \tau/m}$.


One may divide the inverse temperature $\beta$ into $M$ smaller temperatures
$\tau=\beta/M$ and label each temperature by $n=1,2,\ldots,M$,
with the configuration at temperature $n$ being the $6N$-dimensional vector
$\{{\bf p}^{(n)},{\bf q}^{(n)}\}$.
Using the above, the Maxwell-Boltzmann operator
may be written as the successive products of integrals over
these configurations.
Thus the Wigner-Kirkwood commutation function may be written
for the case of quantized momentum,
\begin{eqnarray} \label{Eq:ebHW-disc}
\lefteqn{
e^{-\beta {\cal H}({\bf p},{\bf q})}  e^{W({\bf p},{\bf q})}
} \nonumber \\
& = &
V^N \phi_{\bf p}({\bf q})^*\,
e^{-\beta \hat{\cal H}({\bf q})}
\phi_{\bf p}({\bf q})
\nonumber \\ & = &
e^{{\bf p}\cdot{\bf q}/{\rm i}\hbar}
\prod_{n=1}^M
\left[ e^{-\tau\hat {\cal K}({\bf q})} e^{-\tau U({\bf q})} \right]
e^{-{\bf p}\cdot{\bf q}/{\rm i}\hbar}
\nonumber \\ & = &
e^{{\bf p}\cdot{\bf q}/{\rm i}\hbar}
\prod_{n=1}^{M-1}
\left[ e^{-\tau\hat {\cal K}({\bf q})} e^{-\tau U({\bf q})} \right]
\sum_{ {\bf p}^{(M)}}
\frac{1}{V^N} \int_V {\rm d}{\bf q}^{(M)}
\nonumber \\ & & \mbox{ } \times
e^{-{\bf p}^{(M)}\cdot {\bf q}^{(M,0)}/{\rm i}\hbar}
e^{-\tau {\cal K}({\bf p}^{(M)})}
e^{-\tau U({\bf q}^{(M)})}
e^{-{\bf p}^{(0)}\cdot{\bf q}^{(M)}/{\rm i}\hbar}
\nonumber \\ & = &
\prod_{n=1}^M
\sum_{ {\bf p}^{(n)}}
\frac{1}{V^N} \int_V {\rm d}{\bf q}^{(n)}
e^{-{\bf p}^{(n)}\cdot {\bf q}^{(n,n-1)}/{\rm i}\hbar}
\nonumber \\ & & \mbox{ }\times
e^{-\tau {\cal K}({\bf p}^{(n)})}
e^{-\tau U({\bf q}^{(n)})}
e^{-{\bf p}^{(0)}\cdot{\bf q}^{(M,0)}/{\rm i}\hbar},
\end{eqnarray}
where $ {\bf p}^{(0)} \equiv {\bf p} $, ${\bf q}^{(0)} \equiv {\bf q}$,
and
${\bf q}^{(n,m)} \equiv {\bf q}^{(n)} - {\bf q}^{(m)}$.

Taking the continuum momentum limit yields
\begin{eqnarray} \label{Eq:ebHW}
\lefteqn{
e^{-\beta {\cal H}({\bf p},{\bf q})}  e^{W({\bf p},{\bf q})}
}  \\
& = &
\prod_{n=1}^{M}
\int
\frac{{\rm d}{\bf q}^{(n)}}{\Lambda_\tau^{3N}}
e^{-\pi {q}^{(n,n-1)2}/\Lambda_\tau^2}
e^{-\tau U({\bf q}^{(n)})}
e^{-{\bf p}\cdot{\bf q}^{(M,0)}/{\rm i}\hbar}.\nonumber
\end{eqnarray}
This is the classical phase space weight (continuum momentum)
for a quantum system expressed as a Feynman path integral.
This is  really $N$ ring polymers,
with the beads connected by Gaussians.
Only equivalent beads (ie.\ those with the same $n$)
interact via the potential $U({\bf q}^{(n)})$.
The symmetrization function may simply be factored in.

Integrating over the  momentum continuum
we get the position configuration weight
(without symmetrization)
\begin{eqnarray} \label{Eq:-Bu0+olW}
\lefteqn{
e^{-\beta U({\bf q}^{(0)})} e^{\overline W({\bf q}^{(0)})}
}  \\
& \equiv &
\frac{1}{\Delta_p^{3N}} \int {\rm d}{\bf p}\;
e^{-\beta {\cal H}({\bf p},{\bf q}^{(0)})}  e^{W({\bf p},{\bf q}^{(0)})}
\nonumber \\  & = &
\prod_{n=1}^{M-1}
\frac{1}{\Lambda_\tau^{3N}}
\int {\rm d}{\bf q}^{(n)}\;
e^{-\pi {q}^{(n-1,n)2}/\Lambda_\tau^2}
e^{-\tau U({\bf q}^{(n)})}
\nonumber \\ && \mbox{ } \times
\frac{V^N}{\Lambda_\tau^{3N}}
\int {\rm d}{\bf q}^{(M)}\;
e^{-\pi {q}^{(M-1,M)2}/\Lambda_\tau^2}
e^{-\tau U({\bf q}^{(M)})}
\delta( {\bf q}^{(M,0)})
\nonumber \\  & = &
\frac{ V^N e^{-\tau U({\bf q}^{(0)})} }{\Lambda_\tau^{3NM}}
\int {\rm d}{\bf q}^{(1)}\ldots{\rm d}{\bf q}^{(M-1)}\;
e^{-\pi {q}^{(M-1,0)2}/\Lambda_\tau^2}
\nonumber \\ & & \mbox{ } \times
\prod_{n=1}^{M-1}
\left[
e^{-\pi {q}^{(n-1,n)2}/\Lambda_\tau^2}
e^{-\tau U({\bf q}^{(n)})} \right].
\nonumber \\  & = &\nonumber
\frac{ V^N }{\Lambda_\tau^{3NM}}
\int {\rm d}{\bf q}^{N(M-1)} \;
\prod_{n=1}^{M}
\left[
e^{-\pi {q}^{(n-1,n)2}/\Lambda_\tau^2}
e^{-\tau U({\bf q}^{(n)})} \right],
\end{eqnarray}
with $ {\bf q}^{(M)} \equiv {\bf q}^{(0)}$.
This has the appearance of the usual density matrix (Ceperley 1995).

\subsection{Virial Pressure}

The thermodynamic pressure for a canonical equilibrium system is given
by the volume derivative of the Helmholtz free energy,
which is proportional to the logarithm of the partition function
(Attard 2002, 2012),
\begin{equation} \label{Eq:BpV}
\beta p = \frac{\partial \ln Z}{\partial V}
= \frac{1}{3L^2 Z }  \frac{\partial  Z}{\partial L} .
\end{equation}
The canonical equilibrium partition function
with quantized momentum is formally
\begin{eqnarray} \label{Eq:Znvt}
\lefteqn{
Z(N,V,T)
}  \\
& = &
\frac{1}{N!V^N} \sum_{\bf p}  \int_V {\rm d}{\bf q} \;
e^{-\beta {\cal H}({\bf p},{\bf q})+ W({\bf p},{\bf q})}
\eta({\bf p},{\bf q})
\nonumber \\ & = &
\frac{1}{N!} \sum_{{\bf p}^{(0)}}
\int_V \frac{ {\rm d}{\bf q}^{(0)}  }{V^N}\;
\eta({\bf p}^{(0)},{\bf q}^{(0)})
\nonumber \\ & & \mbox{ }\times
\prod_{n=1}^M
\sum_{ {\bf p}^{(n)}}
\int_V \frac{ {\rm d}{\bf q}^{(n)} }{V^N}
e^{-{\bf p}^{(n)}\cdot {\bf q}^{(n,n-1)}/{\rm i}\hbar}
\nonumber \\ & & \mbox{ }\times
e^{-\tau {\cal K}({\bf p}^{(n)})}
e^{-\tau U({\bf q}^{(n)})}
e^{-{\bf p}^{(0)}\cdot{\bf q}^{(M,0)}/{\rm i}\hbar} .
\nonumber
\end{eqnarray}
The imaginary part of this integrates to zero.

The position integrals are over the cubic volume $V=L^3$.
We scale all lengths by $L$, ${\bf q}^{(n)} = L{\bf x}^{(n)}$,
with $x_{j\alpha}^{(n)} \in [-1/2,1/2]$.
Hence $\partial {\bf q} /\partial L = L^{-1}  {\bf q}$.
The quantized momenta have spacing $\Delta_p = 2\pi\hbar/L$
and so at constant quantum state,
$\partial {\bf p} /\partial L = -L^{-1}  {\bf p}$.
In this case products such as ${\bf p}\cdot {\bf q}$
are independent of $L$.
This includes products with permuted configurations,
${\bf p}\cdot \hat{\rm P} {\bf q}$,
which means that the symmetrization function is independent of $L$,
as well as the Fourier factors that appear in the temperature points.

Since the volume elements are in the form $ {{\rm d}{\bf q}^{(n)} }/{V^N}$,
the only part of the partition function that depends upon $L$
is the kinetic energy,
$\partial {\cal K}({\bf p}^{(n)})/\partial L
= -2L^{-1}{\cal K}({\bf p}^{(n)})$,
and the potential energy, $ \partial U({\bf q}^{(n)}) /\partial L
= L^{-1} {\bf q}^{(n)} \cdot \nabla^{(n)} U({\bf q}^{(n)}) $.
Hence the virial pressure is
\begin{eqnarray} \label{Eq:bp-virial}
\beta p & = &
\frac{\tau}{3V} \sum_{n=1}^M
\left\langle
2  {\cal K}({\bf p}^{(n)})
-  {\bf q}^{(n)} \cdot \nabla^{(n)} U({\bf q}^{(n)})
\right\rangle
\nonumber \\ & = &
\frac{\beta}{3V}  \left\langle
2  {\cal K}({\bf p}^{(1)})
-  {\bf q}^{(1)} \cdot \nabla^{(1)} U({\bf q}^{(1)})
\right\rangle .
\end{eqnarray}
The second equality follows because all temperature point are equivalent.
Having obtained this result for the pressure,
one is now free to take the continuum momentum limit.

This result agrees with Eq.~(6.18) of Ceperley (1995)
(although it depends upon how the kinetic energy is to be calculated
---see below).
However Ceperley (1995 Eq.~(6.17)) also gives
(my notation)
\begin{equation}
p = \frac{1}{3\tau V}
\left\langle
3N
- \frac{\pi {q}^{(n,n-1)2} }{ \Lambda_\tau^2}
+ 2 \tau {\bf q}^{(n)} \cdot \nabla^{(n)} {U}
\right\rangle .
\end{equation}
(I believe that the final term contains a typographical error,
and that one should replace $+2 \Rightarrow -1$.)
For free diffusion  one would expect
$\left({\bf q}^{(n)} - {\bf q}^{(n-1)} \right)^2
= {\cal O}(3N\Lambda_\tau^2)$.
Hence  the first two terms, each of order $NM^{0}$,
have to cancel each other
to leave a residual ${\cal O}(N M^{-1})$,
which in turn converts the prefactor $\tau^{-1} \Rightarrow \beta^{-1}$,
which is essential on physical grounds.
Such exact cancelation of two large statistical averages
is numerically demanding.


\comment{ 
One of the advantages of the present path integral formulation
of the Wigner-Kirkwood commutation function
over the usual path integral formulation of quantum statistical mechanics
(Allen and Tildesley  1987, Ceperley 1995, Feynman and Hibbs 1965)
is that the momentum ${\bf p}$ appears explicitly
in the classical phase space weight,
Eq.~(\ref{Eq:ebHW}).
Hence the usual classical expression for the kinetic energy may be used,
$ {\cal K}({\bf p}) = \sum_{j=1}^N p_j^{2}/2m$.
There is not the ambiguity or need for cancelation of large terms
that is present with other approaches (Ceperley 1995 Eq.~(6.9)).
\textbf{Not so clear. See next.}
} 

\subsection{Kinetic Energy}

From Eq.~(\ref{Eq:ebHW-disc})
we can obtain
the average kinetic energy in the continuum momentum limit,
neglecting symmetrization effects.
We have to evaluate integrals of the form
\begin{eqnarray} \label{Eq:Kpjnj}
\lefteqn{
\left\langle {\cal K}({\bf p}^{(1)}) \right\rangle
} \nonumber \\
& = &
\frac{1}{Z V^N} \sum_{{\bf p}^{(1)}} \int {\rm d}{\bf q}^{(1)}\;
e^{-{\bf p}^{(1)}\cdot {\bf q}^{(1,0)}/{\rm i}\hbar}
\nonumber \\ & & \mbox{ }\times
e^{-\tau {\cal K}({\bf p}^{(1)})}
e^{-\tau U({\bf q}^{(1)})}
{\cal K}({\bf p}^{(1)})
\nonumber \\ & = &
\frac{1}{Z h^{3N}}
\int {\rm d}{\bf p}^{(1)}\, {\rm d}{\bf q}^{(1)}\;
e^{-\tau [ {\bf p}^{(1)} + m{\bf q}^{(1,0)} /{\rm i}\hbar\tau ]^2/2m}
\nonumber \\ & & \mbox{ }\times
e^{- {q}^{(1,0)2}  m/2\hbar^2\tau}
e^{-\tau U({\bf q}^{(1)})}
\nonumber \\ & & \mbox{ }\times
\frac{1}{2m}
\left[ {\bf p}^{(1)} + m{\bf q}^{(1,0)} /{\rm i}\hbar\tau
-  m{\bf q}^{(1,0)} /{\rm i}\hbar\tau \right]^2
\nonumber \\ & = &
\frac{1}{Z h^{3N}}
\int   {\rm d}{\bf q}^{(1)}\;
e^{- \pi {q}^{(1,0)2} /\Lambda_\tau^2}
e^{-\tau U({\bf q}^{(1)})}
\nonumber \\ & & \mbox{ }\times
[2\pi m/\tau]^{3N/2} \left\{
\frac{3N}{2\tau} - \frac{m^2{q}^{(1,0)2}}{2m\hbar^2\tau^2}
\right\}
\nonumber \\ & = &
\frac{1}{ Z \Lambda_\tau^{3N/2}}
\int   {\rm d}{\bf q}^{(1)}\;
e^{- \pi {q}^{(1,0)2} /\Lambda_\tau^2}
e^{-\tau U({\bf q}^{(1)})}
\nonumber \\ & & \mbox{ }\times
\left\{ \frac{3N}{2\tau}
- \frac{\pi {q}^{(1,0)2}}{\tau\Lambda^2_\tau} \right\}
\nonumber \\ & = &
\left\langle  \frac{3N}{2\tau}
- \frac{\pi {q}^{(1,0)2}}{\tau\Lambda^2_\tau} \right\rangle .
\end{eqnarray}
In the limit $M \to \infty$,
we can expect the free diffusion limit,
$\langle {q}^{(1,0)2}\rangle = 3N \Lambda^2_\tau/2\pi$,
and so the two terms should substantially cancel
leaving something ${\cal O}((M\tau)^{-1})$.
This almost agrees with  Ceperley (1995 Eq.~(6.9)),
except that he has an additional contribution from the action.

\subsection{Thermodynamic Energy}

For a classical canonical equilibrium system
the inverse temperature derivative of the Helmholtz free energy
gives the average energy (Attard 2002, 2012),
$\partial (\beta F(N,V,T))/\partial \beta
= -\partial \ln Z(N,V,T)/\partial \beta
= \langle {\cal H} \rangle$.
For the present quantum case with quantized momentum
Eq.~(\ref{Eq:Znvt}) gives
\begin{eqnarray}
\frac{-\partial \ln Z(N,V,T)}{\partial \beta}
& = &
\left\langle {\cal H}({\bf \Gamma}) - \dot W({\bf \Gamma}) \right\rangle
\nonumber \\ & = &
\frac{1}{M} \sum_{n=1}^M
\left\langle {\cal H}({\bf \Gamma}^{(n)}) \right\rangle
\nonumber \\ & = &
\left\langle {\cal H}({\bf \Gamma}^{(0)}) \right\rangle,
\end{eqnarray}
where
$\dot W({\bf \Gamma}) \equiv {\partial W({\bf \Gamma})}/{\partial \beta}$
and
${\cal H}({\bf \Gamma}^{(n)}) =
{\cal K}({\bf p}^{(n)}) + U({\bf q}^{(n)})$.
Having obtained this result one is free to take the continuum limit
and to use Eq.~(\ref{Eq:ebHW}) to evaluate the average.

This result may be compared the conventional expression
(Allen and Tildesley 1987 Eq.~(10.23),
Ceperley 1995 Eq.~(6.7))
(my notation)
\begin{equation}
E =
\left\langle
\frac{3N}{2\tau}
- \frac{\pi q^{(n,n-1)2}}{\Lambda_\tau^2\tau}
+ U({\bf q}^{(n)})
\right\rangle .
\end{equation}
In view of the above result for the kinetic energy,
this is the same as my result.

The classical canonical equilibrium heat capacity is
(Attard 2002 Eq.~(4.54))
\begin{eqnarray}
C_V
& = &
\frac{\partial \overline E(N,V,T)}{\partial T}
\nonumber \\ & = &
k_{\rm B}\beta^2 \frac{\partial  \ln Z(N,V,T)}{\partial \beta^2}
\nonumber \\ & = &
k_{\rm B}\beta^2
\left\langle \Delta(E)^2 \right\rangle .
\end{eqnarray}
For the present quantum case with quantized momentum
Eq.~(\ref{Eq:Znvt}) gives
\begin{eqnarray}
\frac{ C_V }{k_{\rm B}}
& = &
\beta^2\left\langle (\dot W - {\cal H})^2 + \ddot W \right\rangle
-
\beta^2\left\langle \dot W - {\cal H} \right\rangle^2
\nonumber \\ & = &
\frac{\beta^2}{M^2} \sum_{n=1}^M \sum_{l=1}^M
\left\langle {\cal H}({\bf \Gamma}^{(n)}) {\cal H}({\bf \Gamma}^{(l)})
\right\rangle
\nonumber \\ && \mbox{ }
- \left[ \frac{\beta}{M} \sum_{n=1}^M
\left\langle {\cal H}({\bf \Gamma}^{(n)})\right\rangle \right]^2 .
\end{eqnarray}
At not too low temperatures the ring polymer is stiff,
and it is reasonable to take
${\cal H}({\bf \Gamma}^{(l)}) \approx {\cal H}({\bf \Gamma}^{(n)})$,
in which case this reduces to
\begin{eqnarray}
\frac{ C_V }{k_{\rm B}}
& \approx &
\frac{\beta^2}{M} \sum_{n=1}^M
\left\langle {\cal H}({\bf \Gamma}^{(n)})^2 \right\rangle
- \left\langle {\cal H}({\bf \Gamma}^{(0)})\right\rangle^2
\nonumber \\ & = &
\beta^2\left\langle {\cal H}({\bf \Gamma}^{(0)})^2 \right\rangle
- \beta^2\left\langle {\cal H}({\bf \Gamma}^{(0)})\right\rangle^2 .
\end{eqnarray}
In this approximation the continuum momentum approximation
can readily be applied.

%
\section{Gaussian Weight} \label{Sec:Gauss}
\setcounter{equation}{0} \setcounter{subsubsection}{0}
\renewcommand{\theequation}{\arabic{section}.\arabic{equation}}
%

\subsection{Free Ring Walk on a Lattice}

Consecutive configurations on the Feynman path integral
are connected by Gaussian bonds but not by potential energy interactions.
This allows such configurations to overlap,
and the ring polymer to intersect itself.
Thus as a model for the Feynman path
we can consider a  random free ring walk
on a lattice with no prohibition on intersections or back-tracks.
As a consequence, we need only analyze a one-dimensional lattice,
since such an ideal walk in multiple dimensions
is simply the product of these.

Consider an intersecting ring polymer of $M$ atoms
on a one-dimensional  lattice with $M_+=M_-=M/2$,
where the subscript denotes a forward or backward step.
The atoms are ideal and experience no external or interaction energy.
The atom $n$ is preceded by $n_\pm$ steps of each type,
so that $n=n_++n_-$.
The number of ways that it can be at site $l=n_+-n_- < n$
(ie.\ $n_\pm = (n \pm l)/2$,
with $n \pm l$ necessarily even) is
\begin{eqnarray}
w_n(l)
& = &
\frac{n!}{n_+!(n-n_+)!}
\frac{(M-n)!}{(M/2-n_+)!(M/2-n+n_+)!}
\nonumber \\ & = &
\frac{n!}{((n+l)/2)!((n-l)/2)!}
 \\ &  & \mbox{ } \times\nonumber
\frac{(M-n)!}{(M/2-(n+l)/2)!(M/2-((n-l)/2))!} .
\end{eqnarray}
Applying Stirling's approximation
and extracting the $l$-dependence gives
\begin{eqnarray}
\lefteqn{
\ln w_n(l)
} \nonumber \\
& = &
n \ln n - n
+ (M-n) \ln (M-n) - (M-n)
\nonumber \\ &  & \mbox{ }
- ((n+l)/2) \ln((n+l)/2) + ((n+l)/2)
\nonumber \\ &  & \mbox{ }
- ((n-l)/2) \ln((n-l)/2) + ((n-l)/2)
\nonumber \\ &  & \mbox{ }
- ((M-n-l)/2) \ln((M-n-l)/2)
\nonumber \\ &  & \mbox{ }
- ((M-n+l)/2) \ln((M-n+l)/2)
\nonumber \\ &  & \mbox{ }
+ ((M-n-l)/2)
+ ((M-n+l)/2)
\nonumber \\ & \sim &
\frac{-M l^2}{2n(M-n)} .
\end{eqnarray}
Hence the probability that atom $n$ is at site $l$is
\begin{equation}
\wp(l|n) =
\frac{ e^{-M l^2/2n(M-n)}}{\sqrt{2\pi n(M-n)/M}} .
\end{equation}
This takes the continuum limit and assumes unrestricted $l$,
although the combinatorial analysis
is strictly valid for $ l \le \min(n,M-n)$
and $l$ with the same parity as $n$.

The mean square displacement averaged over all atoms in the ring is
\begin{eqnarray}
\langle l^2 \rangle
& = &
\frac{1}{M} \sum_{n=1}^M
\int {\rm d}l\; \wp(l|n)  l^2
\nonumber \\ & = &
\frac{1}{M} \sum_{n=1}^M
\int {\rm d}l\;
\frac{ e^{-M l^2/2n(M-n)} }{\sqrt{2\pi n(M-n)/M}} \; l^2
\nonumber \\ & = &
\frac{1}{M} \sum_{n=1}^M \frac{n(M-n)}{M}
\nonumber \\ & \sim &
\frac{M}{2} -  \frac{M}{3}
= \frac{M}{6} , \quad M \to \infty.
\end{eqnarray}
This says that the density at $l$,
which is proportional to the probability of finding an atom at $l$,
is
\begin{equation}
\rho(l|M)
=
\frac{M}{\sqrt{2\pi M/r }} e^{- r l^2/2M } ,
\quad r=6.
\end{equation}
This is normalized to $M$, the number of atoms on the ring polymer.
The lattice has unit spacing between sites.

To connect this with the Feynman path integral in the continuum
we use for the lattice spacing
the square root of the variance of the original Gaussian
for each step,
$\Lambda_\tau^2/2\pi = \Lambda_\beta^2 /2\pi M$.
Hence the  probability of visiting a configuration
in the neighborhood in the position continuum is
\begin{equation}
\wp({\bf q}'|{\bf q})
=
\frac{1}{\sqrt{\Lambda_\beta^2/r }}
e^{- 2 \pi r ({\bf q}'-{\bf q})^2/2\Lambda_\beta^2} ,
\quad r=6.
\end{equation}
Notice that this is independent of $M$ the number of steps.
One sees that the diffusion of the free ring walk
is about the same as that of a free random walk
with one sixth the number of atoms.

It should come as no surprise that the probability density is Gaussian.
This is a  consequence of the central limit theorem.

This probability is essentially the density of points sampled
by the Feynman path integral.
In the analysis that follows we add to this
an off-set that arises from the interaction potential.
It gives the most likely configuration visited by the path,
which can be different to the starting configuration ${\bf q}$.
We determine the value of the off-set
from the Wigner-Kirkwood commutation function.

\subsection{Biassed Sampling of the Neighborhood} \label{Sec:bias}

In Eq.~(\ref{Eq:-Bu0+olW})
we gave the position configuration weight
that comes from the Wigner-Kirkwood commutation function.
Here we write the exponent as
$\overline U({\bf q}) \equiv
 U({\bf q}) - \beta^{-1} \overline W({\bf q}) $.
We now use the known (Attard 2021 Ch.~8)
high temperature expansion for $W$
to confirm the variance of the Gaussian probability
and to obtain the off-set.

The original Feynman path averaged the exponential of
$M^{-1} \sum_{n=1}^M  \beta U({\bf q}^{(n)})$.
Thus here we average the exponential of $ \beta U({\bf q}')$
over the Gaussian-weighted neighborhood of ${\bf q}$.
This gives the position configuration weight
as the Gaussian average of the ordinary Maxwell-Boltzmann weight,
\begin{eqnarray}
\lefteqn{
e^{-\beta \overline U({\bf q})}
\equiv
e^{-\beta U({\bf q}) + \overline W({\bf q})}
}  \\
& = &
\frac{1}{(\Lambda_\beta^2/r)^{3N/2}}
\int {\rm d}{\bf q}'\;
e^{- 2 \pi r ({\bf q}'-{\bf q}- \overline{\bf \Delta}_{\bf q} )^2
/2 \Lambda_\beta^2}
e^{-\beta U({\bf q}')} .\nonumber
\end{eqnarray}
This is the major result of this paper.

We need to determine the offset $\overline{\bf \Delta}_{\bf q}$,
which biases the sampling away from the initial point ${\bf q}$.
This arises from the interaction potential,
and it is absent from the lattice analysis
of a free ring walk in one dimension.
It partially accounts for the internal interactions
of the configurations at each node of the ring polymer.
The bias does not shift the center of mass of the sampled configurations.
It is necessary to  have such a ${\bf q}$-dependent off-set,
otherwise the Gaussian integral over ${\bf q}$ would be trivial.
Also, the configurations ${\bf q}$ and ${\bf q}'$
enter asymmetrically, unlike the configurations of the ring polymer
for the Feynman path integral.

In the high temperature limit, $\beta \to 0$, $\Lambda_\beta \to 0$,
the Guassian becomes sharply peaked.
To evaluate the integral that gives the weight
of the position configuration we expand about
${\bf q}' = {\bf q} + \overline{\bf \Delta}_{\bf q}$.
With the odd terms vanishing,
the second order Taylor expansion is
\begin{eqnarray}
\lefteqn{
e^{-\beta \overline U({\bf q})}
} \nonumber \\
& = &
\frac{1}{(\Lambda_\beta^2/r)^{3N/2}}
\int {\rm d}{\bf q}'\;
e^{- 2 \pi r \big({\bf q}' - {\bf q} - \overline{\bf \Delta}_{\bf q} \big)^2
/2 \Lambda_\beta^2}
e^{-\beta U({\bf q}')}
\nonumber \\ & = &
\frac{
e^{-\beta U({\bf q} + \overline{\bf \Delta}_{\bf q})}
}{ (\Lambda_\beta^2/r)^{3N/2} }
\int {\rm d}{\bf q}'\;
e^{- 2 \pi r \big({\bf q}' - {\bf q} - \overline{\bf \Delta}_{\bf q} \big)^2
/2 \Lambda_\beta^2}
\left\{ \rule{0cm}{0.5cm}
1
\right. \nonumber \\ && \left. \quad
+
\frac{1}{2} \big({\bf q}' - {\bf q} - \overline{\bf \Delta}_{\bf q} \big)^2
: \left[ \beta^2 \nabla U \nabla U - \beta \nabla \nabla U \right]
_{{\bf q} + \overline{\bf \Delta}_{\bf q}}
\right\}
\nonumber \\ & = &
e^{-\beta U({\bf q} + \overline{\bf \Delta}_{\bf q})}
\left\{ 1
+
\frac{\Lambda_\beta^2}{4\pi r}
\left[ \beta^2 \nabla U \cdot \nabla U - \beta \nabla^2 U \right]
_{{\bf q} + \overline{\bf \Delta}_{\bf q}}
\right. \nonumber \\ && \left. \quad
+ {\cal O}(\Lambda_\beta^4)
\right\} .
\end{eqnarray}
To linear order in $\overline{\bf \Delta}_{\bf q}$ and $\Lambda_\beta^2$,
and setting $r=6$,
this becomes
\begin{eqnarray}
e^{-\beta \overline U({\bf q})}
& = &
e^{-\beta U({\bf q})}
\left\{ 1
- \beta \overline{\bf \Delta}_{\bf q} \cdot \nabla U
-   \frac{\beta\Lambda_\beta^2}{24\pi} \nabla^2 U
\right. \nonumber \\ && \mbox{ } \left.
+   \frac{\beta^2\Lambda_\beta^2}{24\pi} \nabla U \cdot \nabla U
\right\} .
\end{eqnarray}

The high temperature expansions 1 and 2 of the  Wigner-Kirkwood
commutation function give the first few terms as
(Attard 2021 Eqs~(8.13) and (8.55))
\begin{equation}
e^{-\beta \overline U({\bf q})}
=
e^{-\beta U({\bf q})}
\left\{ 1 -\frac{ \beta \Lambda_\beta^2  }{24\pi} \nabla^2 U
+ \frac{\beta^2 \Lambda_\beta^2 }{48\pi} \nabla U \cdot \nabla U
\right\}.
\end{equation}
The coefficient of $\nabla^2 U$ agrees with the above,
which confirms that $r=6$ for the variance.
This is an important affirmation of the validity of the approach.
From the remaining term we obtain 
\begin{equation} \label{Eq:oldelq}
\overline{\bf \Delta}_{\bf q}
=
 \frac{\beta\Lambda_\beta^2 }{48\pi} \nabla U({\bf q})  .
\end{equation}
The contribution of the associated term
to the weight $e^{-\beta\overline U({\bf q})}$ is third order in $\beta$.
That it is divided by $48\pi$ suggests that it and other corrections
will make only a small contribution.
The above expression for $e^{-\beta\overline U({\bf q})}$
is not restricted to high temperatures,
although it remains to establish its performance at any temperature
(see the results in \S\ref{Sec:Results}).

The consequence of this result for the off-set
is that sampling must be biased toward \emph{higher} potential energy,
$U({\bf q}') > U({\bf q})$?!
Although counterintuitive,
this may be interpreted as avoiding the  minima in the potential energy,
which is a little like the requirement for the zero point energy.
Also, the individual Maxwell-Boltzmann factors for each temperature point
of the ring polymer that is the Feynman path integral
have inverse temperature $\tau = \beta/M \to 0$,
which means that entropy is dominant
and that the configuration spends less time at its potential minimum
than with the actual inverse temperature $\beta$.

\subsection{Virial Pressure} \label{Sec:VirP}

The canonical equilibrium partition function,
neglecting the symmetrization function, is
\begin{eqnarray} 
\lefteqn{
Z(N,V,T)
}  \\
& = &
\frac{1}{N!\Lambda_\beta^{3N}}
 \int_V  {\rm d}{\bf q} \;
e^{-\beta \overline U({\bf q})}
\nonumber \\ & = &
\frac{r^{3N/2}}{N!\Lambda_\beta^{6N} }
 \int_V  {\rm d}{\bf q} \, {\rm d}{\bf q}'\;
e^{- 2 \pi r ({\bf q}'-{\bf q}- \overline{\bf \Delta}_{\bf q} )^2
/2 \Lambda_\beta^2}
e^{-\beta U({\bf q}')} .\nonumber
\end{eqnarray}

The pressure is the volume derivative of this,
Eq.~(\ref{Eq:BpV}),
$\beta p = {\partial \ln Z}/{\partial V}
= ({1}/{3L^2 Z })  {\partial  Z}/{\partial L}$.
Again we scale all lengths by $L$.
An ideal part comes from the $6N$-dimensional volume element
${\rm d}{\bf q} \, {\rm d}{\bf q}'\;
= L^{6N} {\rm d}{\bf x} \, {\rm d}{\bf x}'$, namely
\begin{equation}
\beta p^{\rm qu, id}
=
\frac{2N}{V}.
\end{equation}
This is twice the classical ideal gas pressure,
and so we expect some cancelation with the remaining part.
The dimensionless virial can be defined
by the derivative of the exponent of the weight,
\begin{eqnarray}
{\cal V}({\bf q}',{\bf q})
& = &
L \frac{\partial }{\partial L}
\left\{
\frac{- 2 \pi r}{2 \Lambda_\beta^2}
({\bf q}'-{\bf q}- \overline{\bf \Delta}_{\bf q} )^2
-\beta U({\bf q}')
\right\}
\nonumber \\ & = &
\frac{- 2 \pi r}{\Lambda_\beta^2}
\big({\bf q}'-{\bf q}- \overline{\bf \Delta}_{\bf q} \big) \cdot
\left({\bf q}'-{\bf q}
- L \frac{\partial \overline{\bf \Delta}_{\bf q} }{\partial L}
\right)
 \nonumber \\ &&  \mbox{ }
 - \beta {\bf q}' \cdot \nabla U({\bf q'}) ,
\end{eqnarray}
with the derivative of the off-set being
\begin{eqnarray}
L \frac{\partial \overline{\bf \Delta}_{\bf q} }{\partial L}
& = &
\frac{\beta\Lambda_\beta^2 }{48\pi}
L \frac{\partial   }{\partial L}
\left\{ L^{-1} \nabla_{\bf x} U(L{\bf x}) \right\}
\nonumber \\ & = &
\frac{\beta\Lambda_\beta^2 }{48\pi}
\left\{ -\nabla U({\bf q})
+ \nabla [{\bf q} \cdot \nabla U({\bf q}) ]\right\}
\nonumber \\ & = &
\frac{\beta\Lambda_\beta^2 }{48\pi}
{\bf q} \cdot \nabla \nabla U({\bf q}).
\end{eqnarray}
Hence the pressure is given by
\begin{equation}
\beta p
=
\frac{2N}{V}
+ \frac{1}{3V} \big\langle {\cal V}({\bf q}',{\bf q}) \big\rangle .
\end{equation}

For a non-interacting system,
$U=0$,
the virial is
${\cal V}^{\rm id} = ({- 2 \pi r}/{\Lambda_\beta^2}) ({\bf q}'-{\bf q})^2$,
and
\begin{equation}
\big\langle {\cal V}^{\rm id} \big\rangle_{\rm id}
=
\frac{- 2 \pi r}{\Lambda_\beta^2} \frac{3N  \Lambda_\beta^2}{2\pi r}
=
- 3N  .
\end{equation}
This follows because for a homogeneous ideal system
we can set ${\bf q} = {\bf 0}$
and we are left with a $3N$-dimensional Gaussian average of $q'^2$.
In this case the total pressure is $\beta p^{\rm id} = N/V$,
which is the same as the classical ideal gas result.

\subsection{Thermodynamic Energy}

The analogue of the average energy
is given by the negative of the inverse temperature derivative
of the logarithm of the partition function
(Attard 2002, 2012),
\begin{eqnarray}
\lefteqn{
\overline E^{\rm tot} \equiv
\frac{-\partial \ln Z(N,V,T)}{\partial \beta}
} \nonumber \\
& = &
\frac{3N}{\beta}
+
\left\langle U({\bf q}')
-
\frac{2 \pi r}{2\beta\Lambda_\beta^2}
({\bf q}' - {\bf q} - \overline{\bf \Delta}_{\bf q})^2
\right. \nonumber \\ && \left. \mbox{ }
- \frac{2 \pi r}{\Lambda_\beta^2}
({\bf q}' - {\bf q} - \overline{\bf \Delta}_{\bf q})
\cdot  \dot{\overline{\bf \Delta}}_{\bf q}
\right\rangle
\nonumber \\ & \equiv &
\frac{3N}{\beta}
+
\big\langle E({\bf q}',{\bf q}) \big\rangle .
\end{eqnarray}
Here
\begin{equation}
\dot{\overline{\bf \Delta}}_{\bf q}
\equiv
\frac{\partial {\overline{\bf \Delta}}_{\bf q}}{\partial \beta}
=
\frac{2}{\beta} {\overline{\bf \Delta}}_{\bf q} .
\end{equation}

In the non-interacting case that $ U=0$,
this is $\overline E^{\rm id,cl} = 3N/\beta  - 3N/2\beta = 3N/2\beta$,
which is  the average energy of the classical ideal gas.

The classical canonical equilibrium heat capacity is
(Attard 2002 Eq.~(4.54))
\begin{eqnarray}
C_V
& = &
\frac{\partial \overline E^{\rm tot} }{\partial T}
=
-k_{\rm B}\beta^2  \frac{\partial \overline E^{\rm tot} }{\partial \beta}
\nonumber \\ & = &
-k_{\rm B}\beta^2
\left\{
\frac{-3N}{\beta^2}
+
\left\langle \dot E({\bf q}',{\bf q})
- E({\bf q}',{\bf q})^2 \right\rangle
\right. \nonumber \\ && \left. \mbox{ }
+ \big\langle E({\bf q}',{\bf q}) \big\rangle^2
\right\},
\end{eqnarray}
where
\begin{eqnarray}
\lefteqn{
\dot E({\bf q}',{\bf q})
 \equiv
\frac{\partial E({\bf q}',{\bf q})}{\partial \beta}
} \nonumber \\
& = &
\frac{\partial }{\partial \beta}
\left\{
U({\bf q}')
-
\frac{2 \pi r}{2\beta\Lambda_\beta^2}
({\bf q}' - {\bf q} - \overline{\bf \Delta}_{\bf q})^2
\right. \nonumber \\ && \left. \mbox{ }
- \frac{2 \pi r}{\Lambda_\beta^2}
({\bf q}' - {\bf q} - \overline{\bf \Delta}_{\bf q})
\cdot  \dot{\overline{\bf \Delta}}_{\bf q}
\right\}
\nonumber \\ & = &
\frac{2\pi r}{\beta^2\Lambda_\beta^2}
({\bf q}' - {\bf q} - \overline{\bf \Delta}_{\bf q})^2
+
\frac{4\pi r}{\beta\Lambda_\beta^2}
({\bf q}' - {\bf q} - \overline{\bf \Delta}_{\bf q})
\cdot \dot{\overline{\bf \Delta}}_{\bf q}
\nonumber \\ && \mbox{ }
+ \frac{2 \pi r}{\Lambda_\beta^2}
\dot{\overline{\bf \Delta}}_{\bf q}
\cdot  \dot{\overline{\bf \Delta}}_{\bf q}
- \frac{2 \pi r}{\Lambda_\beta^2}
({\bf q}' - {\bf q} - \overline{\bf \Delta}_{\bf q})
\cdot  \ddot{\overline{\bf \Delta}}_{\bf q} ,
\end{eqnarray}
and
$ \ddot{\overline{\bf \Delta}}_{\bf q}
\equiv
{\partial \dot{\overline{\bf \Delta}}_{\bf q} }/{\partial \beta}
= {2} {\overline{\bf \Delta}}_{\bf q} /{\beta^2} $.


%
\section{Second Virial Coefficient at High Temperatures} \label{Sec:B2}
\setcounter{equation}{0} \setcounter{subsubsection}{0}
%

In the classical phase space formulation of quantum statistical mechanics
the virial pressure is formally
\begin{equation}
\beta p
=
\frac{1}{3V }
\Big\langle
{\bf \Gamma}^\dag \cdot \nabla_{\bf \Gamma}
\left[-\beta {\cal H}({\bf \Gamma}) + W({\bf \Gamma}) \right]
\Big\rangle .
\end{equation}
Here a point in classical phase space is
${\bf \Gamma} = \{{\bf p},{\bf q}\}$,
its conjugate is ${\bf \Gamma}^\dag = \{-{\bf p},{\bf q}\}$,
and ${\bf \Gamma}^\dag \cdot \nabla_{\bf \Gamma}
= -{\bf p} \cdot \nabla_p + {\bf q} \cdot \nabla_q$.
The classical Hamiltonian consists of kinetic and potential energies,
${\cal H}({\bf \Gamma}) = {\cal K}({\bf p}) + U({\bf q})$.
In what follows we assume that the potential energy
is the sum of central pair interactions,
\begin{equation}
U({\bf q})
=   \sum_{j<k} u(q_{jk})
= \frac{1}{2} \sum_{j,k} u_{jk} .
\end{equation}
Obviously $j \ne k$ here and in similar sums below.
The virial derivative of the Hamiltonian is
\begin{equation}
{\bf \Gamma}^\dag \cdot \nabla_{\bf \Gamma}
[-\beta {\cal H}({\bf \Gamma})  ]
=
\frac{\beta}{m} \sum_j p_j^2
- \frac{\beta}{2} \sum_{j,k} 
\frac{u'_{jk} }{q_{jk}} q_{jk}^2 .
\end{equation}

\subsection{Commutation Function Contribution}  \label{Sec:B2W2}

A variety of expansions in powers of either Planck's constant
or inverse temperature for the Wigner-Kirkwood commutation function  exist
(Attard 2021 Ch.~8).
In what follows we focus on
the leading term in the expansion of the exponent,
which is quadratic in the inverse temperature (Attard 2021 \S8.5)
\begin{equation}
W^{(2)}({\bf \Gamma}) =
\frac{-\beta^2\hbar^2}{4m} \nabla^2 U({\bf q})
-
\frac{\mathrm{i}\beta^2\hbar}{2m} {\bf p} \cdot \nabla  U({\bf q}) .
\end{equation}

We have for the real part
\begin{equation}
W^{(2)}_{\rm r}({\bf q})
 =
\frac{-\beta^2\hbar^2}{4m} \sum_{j,k} \!^{(k \ne j)}
\left[ u''_{jk} + \frac{2u'_{jk}}{q_{jk}} \right] ,
\end{equation}
which has gradient
\begin{equation}
{\bf q} \cdot \nabla_q W^{(2)}_{\rm r}({\bf q})
=
\frac{-\beta^2\hbar^2}{4m} \sum_{j,k} 
 \left[ u'''_{jk}
 + \frac{2u''_{jk}}{q_{jk}}
  - \frac{2u'_{jk}}{q_{jk}^2}  \right] q_{jk} .
\end{equation}
The imaginary part may be written
\begin{eqnarray}
W^{(2)}_{{\rm i} }
& = &
\frac{ - \beta^2\hbar}{2m}
\sum_{j,k} \!^{(j \ne k)}
\frac{u'_{jk}}{q_{jk}} {\bf p}_{j} \cdot {\bf q}_{jk}
\equiv
\frac{-1}{\hbar}  {\bf p}  \cdot  {\bf a}   ,
\nonumber \\ &  & \quad
{\bf a}_{j}   \equiv
\frac{\beta^2\hbar^2}{2m}
\sum_{k}\!^{(k \ne j)} \frac{u'_{jk}}{q_{jk}} {\bf q}_{jk} .
\end{eqnarray}
Hence
\begin{eqnarray}
{\bf q} \cdot \nabla_q W^{(2)}_{\rm i}
& = &
\frac{ -\beta^2\hbar}{4m}
\sum_{j,k}\!^{(k \ne j)}
u''_{jk}  {\bf p}_{jk} \cdot {\bf q}_{jk}
\equiv
 \frac{-1}{\hbar} {\bf p} \cdot {\bf b} ,
\nonumber \\ &  & \quad
{\bf b}_j \equiv
\frac{\beta^2\hbar^2}{2m}  \sum_{k}\!^{(k \ne j)} u''_{jk} {\bf q}_{jk} ,
\end{eqnarray}
and ${\bf p} \cdot \nabla_p W^{(2)}_{\rm i}
= -{\bf p} \cdot {\bf a}/{\hbar} $.

We have several momentum integrals to evaluate.
The first is
\begin{eqnarray}
\lefteqn{
\frac{1}{h^{3N}} \int {\rm d}{\bf p}\;
e^{-\beta {\cal K}({\bf p})}
e^{ {\rm i} W^{(2)}_{\rm i}({\bf \Gamma}) }
} \nonumber \\
& = &
\frac{1}{h^{3N}} \int {\rm d}{\bf p}\;
e^{-\beta p^2/2m} e^{-{\rm i} {\bf p} \cdot {\bf a}/\hbar}
\nonumber \\ & = &
\frac{1}{h^{3N}} \int {\rm d}{\bf p}\;
e^{-\beta [{\bf p} + {\rm i}m{\bf a} /\beta\hbar]^2/2m}
e^{-  m a^2 /2\beta\hbar^2}
\nonumber \\ & = &
\frac{ e^{- 2\pi a^2 /2\Lambda_\beta^2} }{ \Lambda_\beta^{3N} }.
\end{eqnarray}
The second is
\begin{eqnarray}
\lefteqn{
\frac{1}{h^{3N}} \int {\rm d}{\bf p}\;
e^{-\beta {\cal K}({\bf p})}
e^{ {\rm i} W^{(2)}_{\rm i}({\bf \Gamma}) }
[ -{\rm i}{\bf p} \cdot \nabla_p W^{(2)}_{\rm i} ]
} \nonumber \\
& = &
\frac{1}{h^{3N}} \int {\rm d}{\bf p}\;
e^{-\beta p^2/2m} e^{-{\rm i} {\bf p} \cdot {\bf a}/\hbar}
\frac{{\rm i}}{\hbar} {\bf p} \cdot {\bf a}
\nonumber \\ & = &
\frac{1}{h^{3N}} \int {\rm d}{\bf p}\;
e^{-\beta [{\bf p} + {\rm i}m{\bf a} /\beta\hbar]^2/2m}
e^{-  m a^2 /2\beta\hbar^2}
\nonumber \\ && \quad \times
\frac{{\rm i}}{\hbar} \left[ {\bf p} + \frac{{\rm i}m}{\beta\hbar} {\bf a}
- \frac{{\rm i}m}{\beta\hbar}{\bf a} \right] \cdot {\bf a}
\nonumber \\ & = &
\frac{m a^2}{\beta\hbar^2}
\frac{ e^{- 2\pi a^2 /2\Lambda_\beta^2} }{ \Lambda_\beta^{3N} } .
\end{eqnarray}
The third is
\begin{eqnarray}
\lefteqn{
\frac{1}{h^{3N}} \int {\rm d}{\bf p}\;
e^{-\beta {\cal K}({\bf p})}
e^{ {\rm i} W^{(2)}_{\rm i}({\bf \Gamma}) }
  2 \beta  {\cal K}({\bf p})
} \nonumber \\
& = &
\frac{2\beta}{2m h^{3N}} \int {\rm d}{\bf p}\;
e^{-\beta [{\bf p} + {\rm i}m{\bf a} /\beta\hbar]^2/2m}
e^{-  m a^2 /2\beta\hbar^2}
\; p^2
\nonumber \\ & = &
\frac{2\beta}{2m h^{3N}} \int {\rm d}{\bf p}\;
e^{-\beta [{\bf p} + {\rm i}m{\bf a} /\beta\hbar]^2/2m}
e^{-  m a^2 /2\beta\hbar^2}
\nonumber \\ && \quad \times
\left[{\bf p} + \frac{{\rm i}m{\bf a} }{\beta\hbar}
- \frac{{\rm i}m{\bf a} }{\beta\hbar} \right]^2
\nonumber \\ & = &
\left[ 3 N - \frac{m a^2 }{\beta\hbar^2} \right]
\frac{ e^{- 2\pi a^2 /2\Lambda_\beta^2} }{ \Lambda_\beta^{3N} }.
\end{eqnarray}

With the first of these and neglecting the symmetrization function,
the partition function reduces to
\begin{eqnarray}
Z
& = &
\frac{1}{N! h^{3N}}
\int  {\rm d}{\bf q}\;
e^{-\beta U({\bf q})}
e^{  W^{(2)}_{\rm r}({\bf q}) }
\int {\rm d}{\bf p}\;
e^{-\beta {\cal K}({\bf p})}
e^{ {\rm i} W^{(2)}_{\rm i}({\bf \Gamma}) }
\nonumber \\ & = &
\frac{1}{ N! \Lambda_\beta^{3N} }
\int {\rm d}{\bf q}\;
e^{-\beta U({\bf q})}
e^{  W^{(2)}_{\rm r}({\bf q}) }
e^{- 2\pi a^2 /2\Lambda_\beta^2} .
\end{eqnarray}
The pressure is (again neglecting the symmetrization function)
\begin{eqnarray}
\beta p
& = &
\frac{1}{3V }
\Big\langle
\left\{ -{\bf p} \cdot \nabla_p + {\bf q} \cdot \nabla_q \right\}
\left[-\beta {\cal H}({\bf \Gamma}) + W({\bf \Gamma}) \right]
\Big\rangle
\nonumber \\ & = &
\frac{1}{3V }
\left\langle
2 \beta {\cal K}
- {\rm i}{\bf p} \cdot \nabla_p W^{(2)}_{\rm i}
+ {\rm i} {\bf q} \cdot \nabla_q W^{(2)}_{\rm i}
\right. \nonumber \\ && \left. \quad
- \beta {\bf q} \cdot \nabla_q U
+ {\bf q} \cdot \nabla_q W^{(2)}_{\rm r}
\right\rangle
\nonumber \\ & = &
\frac{1}{3V }
\left\langle
2 \beta {\cal K}
+ \frac{{\rm i}}{\hbar} {\bf p} \cdot [ {\bf a} - {\bf b} ]
- \beta {\bf q} \cdot \nabla_q U
\right. \nonumber \\ && \left. \quad
+ {\bf q} \cdot \nabla_q W^{(2)}_{\rm r}
\right\rangle
\nonumber \\ & = &
\frac{1}{3V }
\left\langle
3 N - \frac{m a^2}{\beta\hbar^2}
+ \frac{{\rm i}}{\hbar} \frac{-{\rm i}m {\bf a}}{\beta \hbar}
\cdot [ {\bf a} - {\bf b} ]
- \beta {\bf q} \cdot \nabla_q U
\right. \nonumber \\ && \left. \quad
+ {\bf q} \cdot \nabla_q W^{(2)}_{\rm r}
\right\rangle
\nonumber \\ & = &
\frac{1}{3V }
\left\langle
3 N
-  \frac{ m }{\beta \hbar^2} {\bf a} \cdot {\bf b}
- \beta {\bf q} \cdot \nabla_q U
+ {\bf q} \cdot \nabla_q W^{(2)}_{\rm r}
\right\rangle .
\nonumber \\
\end{eqnarray}
This is now an average over position,
with the weight that appears in the integrand of the partition function.
The first term is that of the classical ideal gas,
which gives the first virial coefficient.
The final two terms are pair terms and will contribute directly to the second
virial coefficient.

We need to extract the pair part of ${\bf a}\cdot{\bf b}$ ,
as well as of the exponent of the weight, $a^2$.
For these we get
\begin{eqnarray}
\frac{-\pi }{\Lambda_\beta^2} a^2
& = &
\frac{-\pi }{\Lambda_\beta^2}
\left(\frac{\beta^2\hbar^2}{2m}\right)^2
\sum_j
\sum_{k}\!^{(k \ne j)} \frac{u'_{jk}}{q_{jk}} {\bf q}_{jk}
\nonumber \\ && \quad
\cdot
\sum_{k'}\!^{(k' \ne j)} \frac{u'_{jk'}}{q_{jk'}} {\bf q}_{jk'}
\nonumber \\ & \Rightarrow &
\frac{-\pi }{\Lambda_\beta^2} \frac{\beta^2 \Lambda_\beta^4}{16 \pi^2}
\sum_{j,k}\!^{(k \ne j)} u'^2_{jk}  ,
\end{eqnarray}
and
\begin{eqnarray}
\frac{-m }{\beta\hbar^2} {\bf b} \cdot {\bf a}
& = &
\frac{-m }{\beta\hbar^2}
\left(\frac{\beta^2\hbar^2}{2m}\right)^2
\sum_j
\sum_{k}\!^{(k \ne j)} u''_{jk} {\bf q}_{jk}
\nonumber \\ && \quad
\cdot
\sum_{k'}\!^{(k' \ne j)} \frac{u'_{jk'}}{q_{jk'}} {\bf q}_{jk'}
\nonumber \\ & \Rightarrow &
\frac{-\beta^3\hbar^2}{4m} \sum_{j,k}\!^{(k \ne j)}
u''_{jk}u'_{jk} {q}_{jk} .
\end{eqnarray}
In these
the first equality is a sum of two- and three-body terms,
whereas the second equality is the sum of pair terms only.
The pair terms  dominate at low densities,
and so in the low density limit we have
\begin{eqnarray}
\beta p
& = &
\frac{N}{V }
+
\frac{1}{3V}
\left\langle
\frac{- \beta}{2}  \sum_{j,k} \!^{(k \ne j)} u'_{jk} q_{jk}
\right. \nonumber \\ && \left. \quad
-\frac{\beta^2\hbar^2}{4m} \sum_{j,k} \!^{(k \ne j)}
\left[ u'''_{jk}  + \frac{2u''_{jk}}{q_{jk}}
- \frac{2u'_{jk}}{q_{jk}^2}  \right] q_{jk}
\right. \nonumber \\ && \left. \quad
- \frac{\beta^3\hbar^2}{4m} \sum_{j,k}\!^{(k \ne j)} u''_{jk}u'_{jk} {q}_{jk}
\right\rangle .
\end{eqnarray}

Since the average here is over pair terms alone,
we can define the dimensionless exponent of the weight
of a pair of atoms separated by $r$ as
\begin{equation}
v(r) \equiv
- \beta  u(r)
- \frac{\beta\Lambda_\beta^2}{4\pi} \left[ u''(r) + \frac{2u'(r)}{r} \right]
-  \frac{\beta^2 \Lambda_\beta^2}{8 \pi} u'(r)^2 .
\end{equation}
Similarly the (dimensionless) virial function to be averaged is
\begin{eqnarray}
v^{\rm int}(r)
& \equiv &
-\beta u'(r) r
-\frac{\beta\Lambda_\beta^2}{4\pi}
\left[ u'''(r)r  +  2u''(r) - \frac{2u'(r)}{r}  \right]
\nonumber \\ &&  \quad
- \frac{\beta^2\Lambda_\beta^2}{4\pi} u''(r)u'(r) r .
\end{eqnarray}
The first term is the classical contribution,
and the terms proportional to $\Lambda_\beta^2$
are the leading quantum correction to the second virial coefficient
(neglecting symmetrization effects).
With these the second virial coefficient is
\begin{eqnarray}
B_2 = \frac{2\pi}{3}
\int_0^\infty {\rm d}r\; r^2 \, e^{v(r) } v^{\rm int}(r) .
\end{eqnarray}

\subsection{Symmetrization Function Contribution} \label{Sec:B2sym}

In general the virial derivative of the symmetrization function vanishes,
${\bf \Gamma}^\dag \cdot \nabla_{\bf \Gamma} \eta({\bf \Gamma}) = 0$.
At low densities the leading contribution
to the symmetrization function comes
from separate pair transpositions (ie.\ dimer loops)
and their products,
which can be resummed to give the dimer loop grand potential
(Attard 2018b, 2021, 2025a \S7.3.3).
The corresponding pressure for bosons is given by
\begin{equation}
\beta p^{(2)}
= \frac{-\beta}{V} \Omega^{(2)}
=
\frac{1}{V}
\left\langle \frac{N!}{2\,(N-2)!}
\eta^{(2)}({\bf \Gamma}_a, {\bf \Gamma}_b) \right\rangle ,
\end{equation}
with the dimer symmetrization function being
\begin{equation}
\eta^{(2)}({\bf \Gamma}_a, {\bf \Gamma}_b)
=
e^{- {\bf q}_{ab} \cdot {\bf p}_{sb}/{\rm i}\hbar } .
\end{equation}
The pair of atoms chosen for the average is arbitrary.

The leading term in the high temperature expansion
of the Wigner-Kirkwood commutation function
was given in the preceding subsection.
For two particles it is
\begin{equation}
W^{(2)}_{ab} 
=
\frac{ -{\rm i} \beta^2\hbar}{2m}
\frac{u'_{ab}}{q_{ab}} {\bf p}_{ab} \cdot {\bf q}_{ab}
-
\frac{\beta^2\hbar^2}{2m}
\left[ u''_{ab} + \frac{2u'_{ab}}{q_{ab}} \right] .
\end{equation}
The distinct two-particle density
is normalized to $N(N-1)$,
and in the limit of low density it is
\begin{eqnarray}
\rho^{(2)}({\bf \Gamma}_a,{\bf \Gamma}_b)
& = &
 \sum_{j,k}\!^{(j \ne k)}
\big\langle
\delta({\bf \Gamma}_j-{\bf \Gamma}_a)
\delta({\bf \Gamma}_k-{\bf \Gamma}_b)
\big\rangle
\nonumber \\ & \sim &
\frac{N(N-1)
e^{-\beta [{\bf p}_a^2+{\bf p}_b^2]/2m}
e^{-\beta u(q_{ab})}
e^{W^{(2)}_{ab}}
}{
\int {\rm d}{\bf \Gamma}_a\,{\rm d}{\bf \Gamma}_b\;
e^{-\beta [{\bf p}_a^2+{\bf p}_b^2]/2m}
e^{-\beta u(q_{ab})}
e^{W^{(2)}_{ab}}
}
\nonumber \\ & = &
\frac{\rho^2\Lambda_\beta^{6}}{h^{6}}
e^{-\beta [{\bf p}_a^2+{\bf p}_b^2]/2m}
e^{-\beta u(q_{ab})} e^{W^{(2)}_{ab}}.
\end{eqnarray}
The final equality follows because the integral in the denominator is dominated
by separations beyond the range of the pair potential.
Combining this with the dimer symmetrization function
$\eta^{(2)}({\bf \Gamma}_{ab})$,
the  part of the exponent in the average
that depends on momentum is
\begin{eqnarray}
\lefteqn{
\frac{-\beta}{2m} [{\bf p}_a^2+{\bf p}_b^2]
 - \frac{{\rm i} \beta^2\hbar}{2m}
\frac{u'_{ab}}{q_{ab}} {\bf p}_{ab} \cdot {\bf q}_{ab}
- \frac{ {\bf q}_{ab} \cdot {\bf p}_{ab}}{ {\rm i}\hbar }
} \nonumber \\
& = &
\frac{-\beta}{2m} \left[
{\bf p}_a
+ \frac{{\rm i} \beta\hbar u'_{ab}}{2q_{ab}}{\bf q}_{ab}
- \frac{{\rm i}m}{\beta\hbar} {\bf q}_{ab}
\right]^2
\nonumber \\ && \quad
- \frac{\beta}{2m} \left[ \frac{\beta\hbar u'_{ab}}{2q_{ab}}
- \frac{ m}{\beta\hbar} \right]^2 q_{ab}^2
\nonumber \\ &  &\quad
- \frac{\beta}{2m} \left[
{\bf p}_b
+ \frac{{\rm i} \beta\hbar u'_{ba}}{2q_{ba}}{\bf q}_{ba}
- \frac{{\rm i}m}{\beta\hbar} {\bf q}_{ba}
\right]^2
\nonumber \\ && \quad
- \frac{\beta}{2m} \left[ \frac{\beta\hbar u'_{ba}}{2q_{ba}}
- \frac{ m}{\beta\hbar} \right]^2 q_{ba}^2 .
\end{eqnarray}
With this the momentum integrals are readily performed,
and the low density limit of the dimer pressure is
\begin{eqnarray}
\lefteqn{
\beta p^{(2)}
} \nonumber \\
& = &
\frac{1}{2V}
\int {\rm d}{\bf \Gamma}_a\,{\rm d}{\bf \Gamma}_b\;
\rho^{(2)}({\bf \Gamma}_a,{\bf \Gamma}_b)\;
\eta^{(2)}({\bf \Gamma}_a, {\bf \Gamma}_b)
\nonumber \\ & \sim &
\frac{\rho^2\Lambda_\beta^{6}}{2Vh^{6}}
\int {\rm d}{\bf \Gamma}_a\,{\rm d}{\bf \Gamma}_b\;
e^{-\beta [{\bf p}_a^2+{\bf p}_b^2]/2m}
e^{-\beta u(q_{ab})} e^{W^{(2)}_{ab}}
\nonumber \\ && \quad \times
e^{- {\bf q}_{ab} \cdot {\bf p}_{ab}/{\rm i}\hbar }
\nonumber \\ & = &
\frac{\rho^2}{2V}
\int {\rm d}{\bf q}_a\,{\rm d}{\bf q}_b\;
e^{-\beta u(q_{ab})}
e^{ (-{m}/{\beta \hbar^2}) \left[
1 - {\beta^2\hbar^2 u'_{ab}}/{2mq_{ab}} \right]^2 q_{ab}^2 }
\nonumber \\ && \quad \times
e^{(-{\beta^2\hbar^2}/{2m})[ u''_{ab} + {2u'_{ab}}/{q_{ab}}] }
\nonumber \\ & = &
\frac{4\pi\rho^2}{2}
\int_0^\infty {\rm d}r \, r^2\;
e^{-\beta u(r)}
e^{ 
- 4\pi
\left[
1 - {\beta^2\hbar^2 u'(r)}/{2mr} \right]^2 r^2 /2\Lambda_\beta^2}
\nonumber \\ && \quad \times
e^{-{\beta^2\hbar^2}[ u''(r) + {2u'(r)}/{r}]/{2m} } .
\end{eqnarray}
This one-dimensional quadrature is readily performed numerically.
The second virial coefficient due to symmetrization is
$B_2^{\rm sym} = \beta p^{(2)} /\rho^2$.

At high temperatures $\Lambda_\beta \to 0 $,
and the middle exponential is zero unless
$r\alt \Lambda_\beta /\sqrt{2\pi} $.
But if this is in this core region,
the first exponential is zero
(because $u(r) \to \infty$ for $r \alt \sigma$).
Hence one sees that at  low densities,
with decreasing temperature
the contribution to the pressure from symmetrization effects
remains negligible
until the temperature is low enough that
the thermal wavelength is on the order of  the core diameter,
 $\Lambda_\beta \agt \sigma $.

%
\section{Computational Results} \label{Sec:Results}
\setcounter{equation}{0} \setcounter{subsubsection}{0}
\renewcommand{\theequation}{\arabic{section}.\arabic{equation}}
%

\subsection{Algorithm}

For the Gaussian sampling quantum Monte Carlo,
\S~\ref{Sec:Gauss},
the partition function and averages
require a $6N$-dimensional integral
over ${\bf q}$ and  ${\bf q}'$.
Thus it has about the same computational burden
as a path integral simulation with $M=2$,
which is about twice the cost of a classical simulation
with the same number of atoms.
In the following results $N = 1,000$.
The standard Metropolis Monte Carlo algorithm
(Allen and Tildesley 1987, Attard 2002, 2012)
in position configuration space
was adapted to the present problem in an obvious fashion.
Trial moves in ${\bf q}_j$ and ${\bf q}_j'$ were made simultaneously.
The acceptance rate was typically 10--60\%.
Averages were accumulated once every 10 cycles of trial moves.
A small cell, position-based neighbor table for ${\bf q}$
was used to calculate both the potential energy $U({\bf q}')$
and its gradient $\nabla U({\bf q})$.
It was found that  $|{\bf q}_j-{\bf q}_j'| \ll \sigma$
($\sigma =0.255$\,nm is the atomic diameter of helium (van Sciver 2012))
for almost all configurations.
This did increase with decreasing temperature,
and at the lowest temperature studied (5\,K)
in some cases reached $\sigma$.
It was almost always the case that for neighboring atoms,
$q_{jk}^2 \approx q_{jk}'^2 \approx {\bf q}_{jk} \cdot {\bf q}_{jk}'
\alt \sigma$.
The first two increased with decreasing temperature.
On occasion at the lowest temperature studied,
${\bf q}_{jk} \cdot {\bf q}_{jk}' < 0$.
A potential cut-off $R_{\rm cut} = 3.5\sigma$ was used,
and a tail correction was added to the energy and the pressure.
Simulations for a thermodynamic state point took typically 2--4 hours
on a desk-top four-processor personal computer.
The statistical error in the pressure was typically less than 1\%.

The Aziz HFD-B2 pair potential (Aziz 1992) was used.
Limited comparison was made with the Lennard-Jones 6-12 pair potential
(van Sciver 2012).

In addition to the Gaussian weighted sampling,
results were obtained for the second virial coefficient
derived from the second order high temperature expansion
of the Wigner-Kirkwood commutation function in classical phase space
(Attard 2018b, 2021).
The analysis is in \S~\ref{Sec:B2}.

\subsection{Results}

\begin{figure}[t]
\centerline{ \resizebox{8cm}{!}{ \includegraphics*{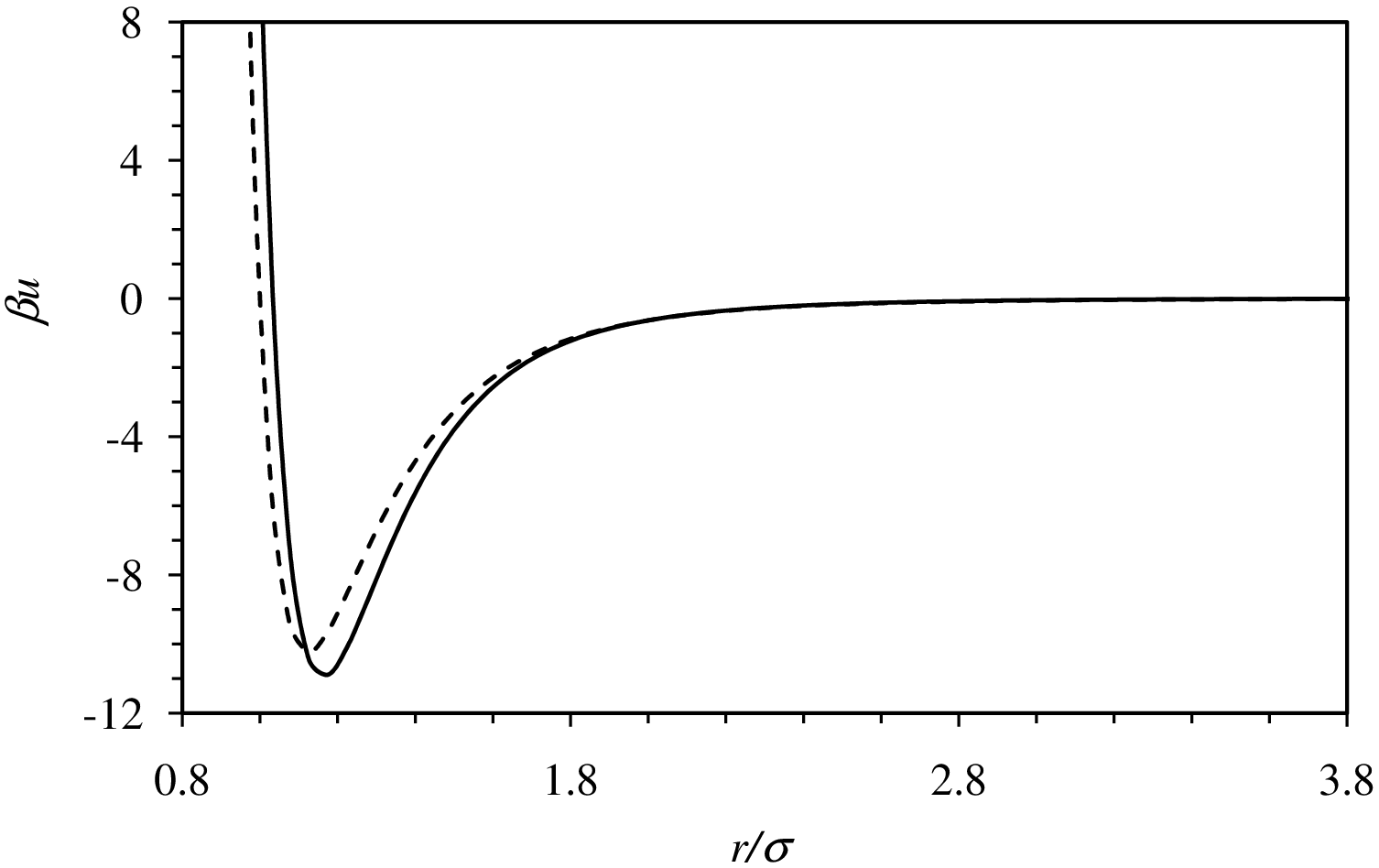} } }
\caption{\label{Fig:u(r)}
Pair potential for $^4$He at $T=1$\,K.
The solid curve is the HFD-B2 potential (Aziz \emph{et al.}\  1992),
and the dashed curve is the Lennard-Jones  potential (van Sciver 2012).
}
\end{figure}

The Lennard-Jones 6-12 pair potential
is perhaps the simplest intermolecular potential
that is capable of describing the three physical phases of matter.
Fitted to the noble gases, the long range $r^{-6}$ attractive tail
is asymptotically exact,
whilst the short-range $r^{-12}$ repulsion is realistic but more approximate.
More reliable pair potentials,
which are necessarily more complicated,
have been obtained from a combination of \emph{ab initio} quantum calculations
and fitting to experimental data (Aziz \emph{et al.}\  1992).
For the noble gases these typically yield agreement
with measured quantities
such as the second virial coefficient or transport coefficient
within the experimental error (Aziz \emph{et al.}\  1992).
One such pair potential, the Hartree-Fock dispersion HFD-B2 potential,
is compared to the Lennard-Jones potential for $^4$He
in Fig.~\ref{Fig:u(r)}.
It can be seen that the two are in remarkably good agreement,
with the HFD-B2 potential having a slightly deeper potential well
that is shifted to a slightly larger separation.
It is debatable whether the greater reliability of the HFD-B2 potential
outweighs its greater complexity and computational burden
compared to the Lennard-Jones potential,
particularly given that three-body and higher order potentials are neglected
in both cases.

\begin{figure}[t]
\centerline{ \resizebox{8cm}{!}{ \includegraphics*{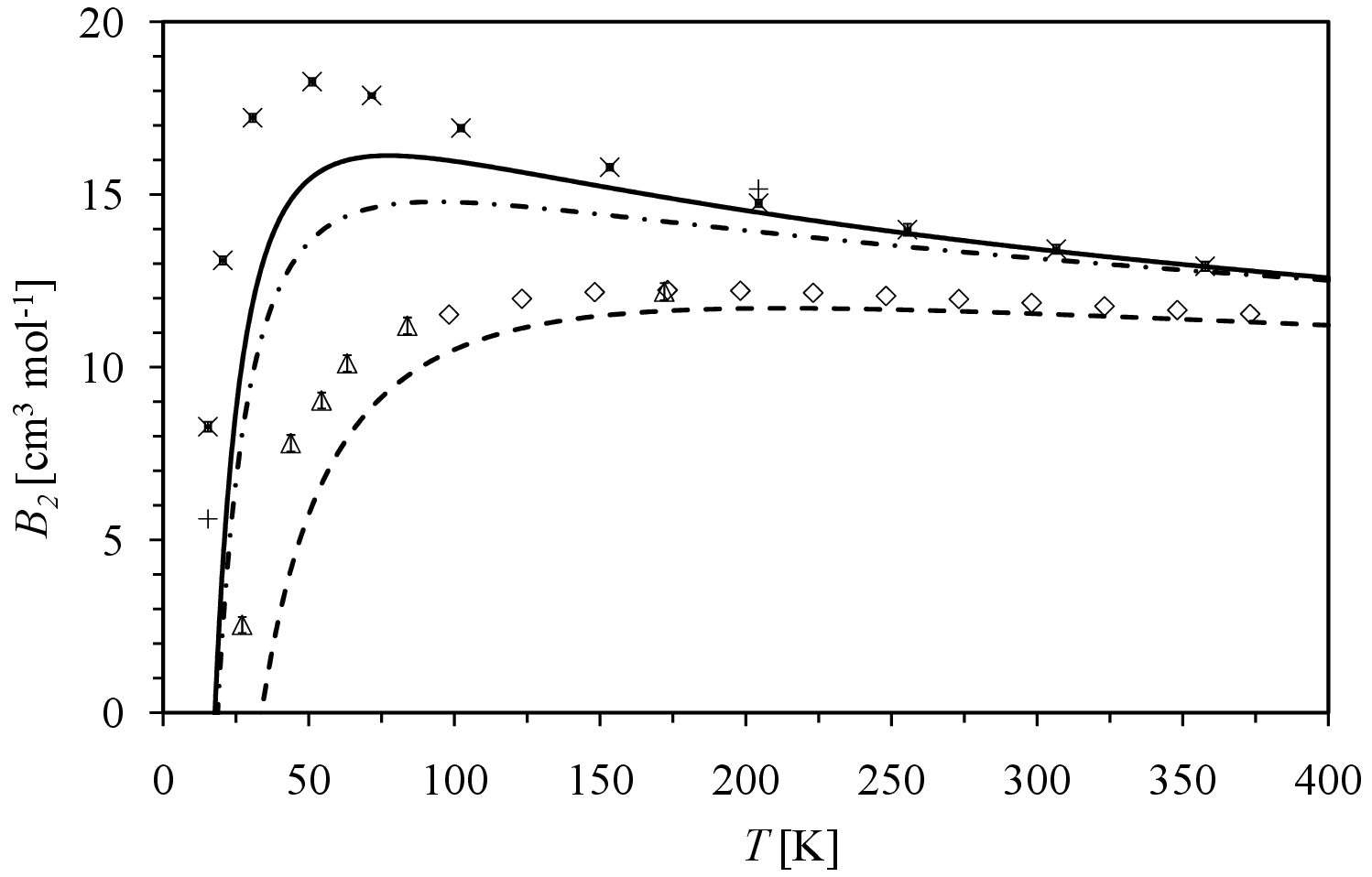} } }
\caption{\label{Fig:B2a}
Second virial coefficient of $^4$He.
The open symbols are measured data
of Gammon (1976) (circles)
and of Kemp \emph{et al.}\ (1986/87) (triangles).
The curves are the analytic second virial coefficient,
\S~\ref{Sec:B2W2},
with the dashed curve being the HFD-B2 classical result, $W^{(2)} = 0$,
the solid curve being the HFD-B2 quantum result,
and the dash-dotted curve being the  Lennard-Jones quantum result.
The quantum Monte Carlo simulations with Gaussian sampling,
\S~\ref{Sec:Gauss},
use the HFD-B2 ($\times$ symbols)
or the Lennard-Jones ($+$ symbols) pair potential.
The symmetrization function is not included in the calculations.
In both the measurements and the simulations
the error is less than the symbol size.
}
\end{figure}

Figure~\ref{Fig:B2a} compares the measured
(Berry  1979, Gammon 1976, Kemp \emph{et al.}\ 1986/87)
and calculated second virial coefficient of $^4$He for $T \agt 20$\,K.
The measurement error is less than the size of the symbols,
and so these can be taken as reliable benchmarks.
In general the calculations tend toward the measured data
as the temperature is increased.
However even at room temperature there is a noticeable
discrepancy between the quantum calculations
and the measured data.
In fact, the classical second virial coefficient
is closer to the measured data than are the present quantum calculations
over most of the temperature range shown.
This is further discussed below

The difference between the calculations with the Lennard-Jones
pair potential and with the HFD-B2 pair potential
are smaller than the difference between the calculations and the measurements.
This confirms the point that at the current stage
of simulations of condensed matter there is an arguable case
for the simpler Lennard-Jones pair potential.

Likewise the difference
between the analytic form for the second virial coefficient,
\S~\ref{Sec:B2W2},
and the quantum Monte Carlo simulations using Gaussian sampling,
\S~\ref{Sec:Gauss},
is relatively minor.
On the one hand this confirms the efficacy of the Gaussian sampling
and of its computational implementation.
On the other hand it emphasizes that it is an approximation
that is better at high temperatures than at low.

An interesting observation
from the quantum Monte Carlo with Gaussian sampling simulations
is that at high temperatures
pairs of $^4$He atoms can approach each other surprisingly closely.
For example, at $T=204$\,K and  $\rho\sigma^3=0.05$,
the minimum separation in a typical configuration of the 1,000 atoms
was 0.76--0.88$\sigma$ for both the Lennard-Jones
and the HFD-B2 pair potentials.
This distance of closest approach increased with decreasing temperature,
as one might expect.
At $T=20.4$\,K and $\rho\sigma^3=0.05$,
it was 0.92--0.98$\sigma$,
and at $\rho\sigma^3=0.02$ it was it was 0.94--1.01$\sigma$.

The symmetrization contribution to the second virial coefficient,
\S~\ref{Sec:B2sym}, is not included in the calculations
in the figure because it is negligible.
At 10\,K, where $\Lambda_\beta = 1.1\sigma$,
the contribution from symmetrization to the second viral coefficient is
about 1/500 the magnitude of the rest.
At 5\,K, where $\Lambda_\beta = 1.5\sigma$, it is about half.

That the second virial coefficient is positive
over the temperature range shown
indicates that it is dominated by the short-range repulsive
part of the pair potential rather than by the attractive tail.
The fact that the classical result is closer to the measured data
than are the two quantum results
indicates that the  leading order high temperature correction,
which has been retained,
must largely cancel with the neglected higher order terms,
and that these are required for full accuracy.
These terms must be strongly influenced by the behavior
of the potential in the repulsive core,
which suggests that the high temperature expansion
is ill-behaved in this region.
This conclusion is supported by other, unpublished,
analysis and calculations by the author,
which include the observation that the higher order gradients
that occur are increasingly divergent in the core.

%
\section{Conclusion} \label{Sec:Concl}
\setcounter{equation}{0} \setcounter{subsubsection}{0}
\renewcommand{\theequation}{\arabic{section}.\arabic{equation}}
%

This paper has replaced the Feynman path integral
that is used in quantum statistical mechanics
by Gaussian sampling in the neighborhood of each point
in position configuration space.
The variance and mean of the Gaussian were determined
from the statistics of a non-interacting ring polymer,
which is what the Feynman path is,
and from the leading term in the high temperature expansion
of the Wigner-Kirkwood commutation function.
A computer algorithm for quantum Monte Carlo based on the Gaussian sampling
was used to obtain the second virial coefficient of $^4$He
for $T \agt 10$\,K.

An analytic formula for the second virial coefficient
at high temperatures was also obtained.
Numerical results were in good agreement
with the Gaussian sampling quantum Monte Carlo results for  $^4$He.
However, they give a second virial coefficient
that is about 10\% larger than the measured value at room temperature,
and about 30--40\% larger at $T=100$\,K.

Undoubtedly the Gaussian sampling could be improved
by going beyond the leading term in the high temperature expansion,
as could the formula for the high temperature second virial coefficient.
However it appears that the existing expansions suffer slow convergence
or even divergence in the core region of the intermolecular potential.
Arguably improving these expansions in this regard
is the most pressing problem.

\section*{References}


\begin{list}{}{\itemindent=-0.5cm \parsep=.5mm \itemsep=.5mm}

\item 
Allen M P and Tildesley D J 1987
\emph{Computer Simulation of Liquids}
(Oxford: Clarendon Press)

\item 
Attard P 2002
\emph{Thermodynamics and Statistical Mechanics:
Equilibrium by Entropy Maximisation}
(London: Academic)

\item 
Attard  P 2012
\emph{Non-equilibrium Thermodynamics and Statistical Mechanics:
Foundations and Applications}
(Oxford: Oxford University Press)

\item 
Attard P 2018b
Quantum statistical mechanics in classical phase space. Expressions for
the multi-particle density, the average energy, and the virial pressure
arXiv:1811.00730

\item 
Attard P  2021
\emph{Quantum Statistical Mechanics in Classical Phase Space}
(Bristol: IOP Publishing)

\item 
Attard P 2025a
\emph{Understanding Bose-Einstein Condensation,
Superfluidity, and High Temperature Superconductivity}
(London: CRC Press)

\item 
Attard P 2025b
The molecular nature of superfluidity: Viscosity of helium from quantum
stochastic molecular dynamics simulations over real trajectories
arXiv:2409.19036v5

\item 
Attard P 2025g
Introduction to the modern theory of Bose-Einstein condensation,
superfluidity, and superconductivity
arXiv:2511.08953

\item   
Aziz R A, Slaman M J,  Koide A, Allnatt  A R, and Meath W J 1992
Exchange-Coulomb potential energy curves for He-He,
and related physical properties
\emph{Mol.\ Phys.}\ {\bf 77} 321

\item
Berry K H 1979
NPL-75: A low temperature gas thermometry scale from 2.6\,K to 27.1\,K
\emph{Metrologia} {\bf 15} 89

\item 
Ceperley  D M  1995
Path integrals in the theory of condensed helium
\emph{Rev.\ Mod.\ Phys.}\ {\bf 67} 279

\item
Feynman R P and Hibbs A R 1965
\emph{Quantum Mechanics and Path Integrals}
(New York: McGraw-Hill)


\item
Gammon B E 1976
The velocity of sound with derived state properties in helium
at -175 to 150 $^\circ$C with pressure to 150~atm.
\emph{J. Chem. Phys.}\ {\bf 64} 2556

\item
Kemp R C, Kemp W R G, and Besley L M 1986/87
A determination of thermodynamic temperatures and measurements
of the second virial coefficient of $^4$He between 13.81\,K and 287\,K
using a constant-volume gas thermometer
\emph{Metrologia} {\bf 23} 61

\item 
Kirkwood J G 1933
Quantum statistics of almost classical particles
\emph{Phys.\ Rev.}\ {\bf 44} 31

\item 
van Sciver  S W 2012
\emph{Helium Cryogenics}
(New York: Springer 2nd edition)

\item
Trotter H F 1959
On  the product of  semi-groups of  operators
\emph{Proc.\ Am.\ Math.\ Soc.}\ {\bf 10} 545

\item 
Wigner E 1932
On the quantum correction for thermodynamic equilibrium
\emph{Phys.\ Rev.}\ {\bf 40} 749

\end{list}



\appendix

%
%

\end{document}